# Neuromimetic Circuits with Synaptic Devices based on Strongly Correlated Electron Systems

Sieu D. Ha[1,†], Jian Shi[1,‡], Yasmine Meroz[1], L. Mahadevan[1,2], and Shriram Ramanathan[1,*]

[1]School of Engineering and Applied Sciences, Harvard University, Cambridge, MA 02138 USA
[2]Department of Physics and Department of Organismic and Evolutionary Biology, Harvard University, Cambridge, MA 02138 USA

[*]Corresponding author: shriram@seas.harvard.edu
[†]Current address: HRL Laboratories LLC, Malibu, CA 90265
[‡]Current address: Department of Materials Science and Engineering, Rensselaer Polytechnic Institute, Troy, NY 12180 USA

**Abstract**

Strongly correlated electron systems such as the rare-earth nickelates ($R$NiO$_3$, $R$ = rare-earth element) can exhibit synapse-like continuous long term potentiation and depression when gated with ionic liquids; exploiting the extreme sensitivity of coupled charge, spin, orbital, and lattice degrees of freedom to stoichiometry. We present experimental real-time, device-level classical conditioning and unlearning using nickelate-based synaptic devices in an electronic circuit compatible with both excitatory and inhibitory neurons. We establish a physical model for the device behavior based on electric-field driven coupled ionic-electronic diffusion that can be utilized for design of more complex systems. We use the model to simulate a variety of associate and non-associative learning mechanisms, as well as a feedforward recurrent network for storing memory. Our circuit intuitively parallels biological neural architectures, and it can be readily generalized to other forms of cellular learning and extinction. The simulation of neural function with electronic device analogues may provide insight into biological processes such as decision making, learning and adaptation, while facilitating advanced parallel information processing in hardware.

# I.  <u>Introduction</u>

In nervous systems, information is processed by a network of neuron cells that each transmit and receive electrical signals through ionic transport.  Neurons are connected through synapses, which control the amount of charge that is transmitted from a preceding (presynaptic) neuron to a subsequent (postsynaptic) neuron, much like resistors in electronic circuits.  Neurons fire signals when the combination of input signals from all preceding neurons is above a threshold value within some time window.  Excitatory (Inhibitory) neurons generate positive (negative) signals that promote (suppress) postsynaptic firing.  A powerful and crucial aspect of nervous systems is that synapse weights can be persistently enhanced or reduced, denoted as plasticity.  This ability is believed to be linked to learning and memory in nervous systems.  At the cellular level, there are several mechanisms by which synapse weight modulation occurs, such as Hebb's Rule, in which temporal correlation between pre- and postsynaptic signals drives weight enhancement.[1]  At the system level, learning often involves many neurons and synapses. Learning generally occurs through associative (e.g. the name of an object becomes attributed its physical appearance) or non-associative (e.g. the startle response to a loud noise reduces with repetition) means.  Simulating neural function with electronics may be a powerful tool for enhancing understanding of system-level brain function and for brain-inspired parallel computation.  However, current software simulations are known to be energy- and space-intensive,[2] and conventional CMOS electronic device implementations are limited by lack of components that mimic biological synapse behavior.[3-5]  Here, we demonstrate a broad array of device-level learning mechanisms using rare-earth nickelate synaptic devices in a circuit architecture analogous to biological systems.  Our nickelate devices and electronic circuit are capable of using both excitatory and inhibitory neurons for associative and non-associative learning, which are some of the core elements in any neural circuit.[6] We start by implementing a Hebbian-like mechanism with excitatory neurons to experimentally demonstrate associative learning. We then show compatibility with inhibitory neurons to experimentally show associative unlearning before generalizing to other learning mechanisms and more complex networks through simulations.

Classical conditioning is a fundamental associative learning process in which an unconditioned stimulus (US) consistently produces an unconditioned response (UR) and an initially neutral stimulus (NS) does not produce a similar response. An example of a US-UR pair in humans is striking of the patellar tendon below the knee and jerking motion of the knee.[7] An NS could be the sound of a bell, which normally does not cause the knee to jerk. In classical conditioning, after sufficient repeated activation of NS directly before US, an association is developed between NS and US such that NS produces a similar response as US, after which the NS is referred to as the conditioned stimulus (CS) and the corresponding response as the conditioned response (CR). Extinguishing of the CS-US association is denoted as unlearning, although there are other processes by which CS stops generating a response, collectively known as extinction.[8]

While classical conditioning often involves a multitude of neuronal connections, individual neurons and synapses can also display conditioning in response to external stimuli.[9] Efforts in constructing electronic conditioning circuits have focused on such cellular-level learning, with one neuron-like component each for the US, CS, and UR/CR (Figure 1a). Fabricating electronic neural circuits with conventional semiconductor technology has not been widespread because of limited options for devices with synaptic characteristics (i.e. modifiable, persistent resistance/charge/spin state). There have been recent experimental demonstrations of conditioning using resistive switches.[10,11]  These devices have two or more stable resistance states between which electronic switching occurs, and they are of interest primarily for computer memory and information processing.[12]  For neural simulation, resistive switches have limitations due to difficulties in obtaining a continuum of stable resistance states,[13] as in biological synapses, with high uniformity between devices.[14]  Moreover, the aforementioned circuit demonstrations with resistive switches, while they do show gradual resistance changes and may be potentially useful for computation,[14-16] are not analogous to biological neural architectures in certain fundamental ways and may not be generalizable to arbitrary neuronal systems.  For example, in the work of Bichler *et al.*,[10] neurons do not explicitly transmit signals through synapses to other neurons, and a non-biological read-write synchronization scheme is necessary to independently modify and measure synaptic weights.  In the work of Ziegler *et al.*,[11] the focus is on demonstrating classical conditioning and not on emulating neural architectures.  Thus, they do not claim to use neuron-like or synapse-like components, and learning functionality is achieved by including plasticity in the signal transmitters (i.e. neurons) themselves, with the requirement that different transmitters need to output dissimilar voltage levels.  While these works show classical conditioning and may have application in computation, our device and resultant network design are more similar to biological neuronal systems, which enables us to demonstrate a greater general variety of neural processes, including associative unlearning, non-associative learning and unlearning, and usage of excitatory and inhibitory neurons.

We have recently shown that 3-terminal electronic devices using the rare-earth nickelate SmNiO$_3$ (SNO) display synaptic behavior when gated with an electrolyte.[17]  SNO has unusual electronic properties such as a temperature-dependent insulator-metal transition arising from strong interactions between electrons, spins, orbitals, and phonons.[18]  These interactions cause the resistivity of SNO to have a much larger range of values as compared to conventional semiconductors for similar changes in defect density,[19] which is highly useful for synapse simulation. In 3-terminal SNO devices (Figure 1c), the room temperature insulating resistance between source and drain terminals can be modulated in a continuous persistent manner by applying a voltage to the gate terminal. In our SNO devices, positive (negative) gate voltage induces resistance increase (decrease). Such plasticity simulates biological long-term potentiation and depression (LTP, LTD), which refer to persistent synaptic weight modification over long time scales relative to signal transmission.[20] As detailed below in Section IV, gating of SNO devices induces reversible electrochemical changes to the oxygen stoichiometry and resistance in the thin films. This can be considered analogous to the biological case, in which synaptic weight modulation involves a change in the number of ion-conducting channels (AMPA-type glutamate receptors) present at the synapse between two neurons.[21] Furthermore, the source and drain terminals of the SNO device are akin to neuronal interfaces between a synapse and its corresponding pre- and postsynaptic neurons. The usage of a third gate terminal for synapse modulation is similar to certain biological learning mechanisms known as retrograde signaling.[22] From an electronic perspective, advantages of 3-terminal devices over 2-terminal resistive switches include reduced sneak path problem (i.e. crosstalk noise),[23] potential low-power operation,[24] and the ability for concurrent signal transmission and LTP/LTD, which is similar to neural circuits and which removes the need for complex circuit timing algorithms.

## II.   Experimental

A detailed description of SmNiO$_3$ growth and device fabrication by lithography can be found elsewhere.[17] SNO films were grown on LaAlO$_3$ (LAO) by co-sputtering Sm and Ni from metallic targets (ATC Orion system from AJA International) followed by annealing in a custom-built furnace at 1500 psi $O_2$ and 500 °C for 24 h.  3-terminal devices were fabricated using standard photolithography (channel dimensions 400 μm x 2000 μm), and the ionic liquid *N,N*-diethyl-*N*-methyl-N-(2-methoxyethyl)ammonium bis(trifluoromethylsulphonyl)imide (DEME-TFSI, Kanto Chemical) was used as the ionic liquid gate dielectric.  SNO devices were interfaced with other circuit components via an environmental probe station (MDC Corporation).  Details of the circuits can be found in Appendix A.  Neuronal stimuli were implemented using Keithley 2400 source-measure units, and neuronal output was monitored with a Keithley 2000 multimeter.  All equipment was controlled with customized LabView code, and all measurements were performed in ambient conditions.  Circuit modeling was implemented using the SimElectronics package in MATLAB/Simulink.  A customized Simscape component was created to model the gated-SNO device behavior.

### III.   Circuit design and experimental learning/unlearning

Our circuit architecture for classical conditioning and unlearning is analogous to that of a biological neural circuit (Figures 1a and 1b, full circuit diagram and explanation in Appendix A). The design is guided by prior software simulations showing conditioning in a circuit with resistive switching devices,[25] which have not yet been realized experimentally. We utilize one neuron-like component each to receive/transmit the US (N1), CS (N2), and UR/CR (N3), and we have one synapse-like block each for the US-UR (Synapse 1) and CS-CR (Synapse 2) connections.  The neurons fire when the respective total input voltage is above a set threshold value.  Each synapse block is composed of an SNO device and a logic block.  The logic block (Figure A1b) temporally correlates forward-propagating signals from the presynaptic neuron with back-propagating signals from the postsynaptic neuron, and it sends the correct signal to the SNO gate corresponding either to LTP (resistance decrease) or LTD (increase). Such temporal correlation occurs in certain classes of Hebbian learning.[25,26] LTP occurs if the signals overlap, and LTD occurs if the presynaptic neuron fires without the postsynaptic neuron.[27] The strength of the logic block design is that it requires no external power supply. The logic block and SNO device together are therefore electrically passive, similar to biological synapses. The input signal arriving at N3 is directly determined by the separate resistances of SNO1 and SNO2. The N1 and N2 voltage signals here are equivalent 2 V square pulses corresponding to excitatory neurons. Our design can be general to any magnitude and polarity input neuron signal. Here, SNO resistance is reversibly modifiable and intuitively related to synaptic weight (conductance ~ weight); synaptic weight is persistent; and no external power is needed for the synapse-like blocks. These are all important characteristics of biological synapses. While we use discrete square voltage pulses as pseudo-envelope functions of neuronal spike trains, all of the functionality should remain for short voltage spikes.

Classical conditioning experimental data are shown in Figure 2a. Initially, we *a priori* set SNO1 to 2.5 kΩ, representing the US, and SNO2 to ~9 kΩ, representing the NS. In the conditioning phase, both N1 and N2 are triggered simultaneously for an extended time period. N3 fires during this phase, which applies a negative voltage to the gate of SNO2 and which creates the association between N2 and N3 through Hebbian learning. After the conditioning phase, probing the inputs shows that both N1 and N2 cause N3 to fire, indicating that the resistance of SNO2 decreased (LTP) during the prior phase in accordance with conditioning (NS → CS). Subsequent measurement of the SNO2 resistance reveals that it decreased to ~3.0 kΩ while that of SNO1 remained relatively unchanged after the conditioning phase. This is clear evidence that our circuit design with SNO synapse-like devices exhibits classical conditioning. After conditioning, N2 firing will now trigger N3 and additional LTP of the SNO2 synaptic connection, regardless of N1. This will occur until SNO2 reaches its minimum resistance value (~1-2 k$\Omega$, corresponding to maximum oxygen content as discussed in section IV). This is similar to the non-associative biological process of sensitization, whereby repeated stimulus causes an enhancement in cellular response.[9,28] The time scale of conditioning is expected to be proportional to $1/A^2$, where $A$ is the SNO channel area. For microscopic devices the time scales should be reduced well into the μs range. For comparison, biological neuronal voltage spikes occur on the ms time scale.[6]

The CS-CR association can be unlearned if the firing of N3 is suppressed when N2 fires. This can be accomplished simply by modifying N1 to simulate an inhibitory neuron (see Appendix A). The remainder of the circuit used for classical conditioning is unchanged. We show experimental unlearning results from this circuit in Figure 2b. Now with N1 as an inhibitory neuron, N3 only fires when N2 is triggered, although both SNO1 and SNO2 are in low resistance states. Note that the input to N1 represents the stimulus, which is still positive, but that the output of N1 is negative. When both N1 and N2 fire, the negative signal from N1 suppresses N3 firing, causing a positive voltage to be applied to the gate of SNO2, increasing the resistance of SNO2 and inducing LTD. After sufficient simultaneous application of N1 and N2, it is shown that N3 no longer responds to N2. Measurement of the resistance of SNO2 before and after the unlearning process showed an increase from ~3.0 k$\Omega$ to ~7.2 k$\Omega$, while that of SNO1 again remained relatively stable. This can be viewed as device-level unlearning. Both the usage of inhibitory neurons and demonstration of unlearning have not been previously shown with electronic devices, but they are crucial components of neural behavior. With one circuit design, we are able to experimentally show classical conditioning, sensitization, and unlearning.

## IV.    SmNiO₃ synaptic modification modeling

To illustrate that the SNO device and our network design are capable of a wide variety of neuronal processes and more complex neural circuits, we model the time and voltage dependencies of potentiation and depression in the SNO devices for use in future designs.[29,30] The resistance modulation in the SNO devices follows an electrochemical mechanism in which oxygen vacancies serve as dopant species to regulate the composition and thereby the resistance of the channel material.[17] The use of the ionic liquid provides electrochemical redox species for reducing and oxidizing the SNO channel. More oxygen vacancies lead to more divalent nickel species, which leads to a higher resistance state of the $SmNiO_{3-x}$ channel. The reverse process oxidizes the divalent nickel species back to trivalent species, recovering the low resistivity state. Thus, the minimum resistance of the synaptic device corresponds to near-ideal maximum oxygen content, and the maximum resistance is relatively unbounded. After a gate voltage is applied to an SNO device to modify the resistance, the resistance may decay slightly towards its pre-gated state when the voltage is removed.[17] However, the resistance will not fully return to its initial state, and the longer the gate voltage is applied, the less decay is observed. This is similar to synapse behavior in biological nervous systems,[6] and it can be qualitatively viewed as the need for repeated conditioning before a behavior is permanently learned.

The rate of divalent nickel species change can be modeled, to first order, by combining the Cottrell equation (time dependence) and Butler-Volmer equation (voltage dependence).[29] Here, the Cottrell equation is applied to model two events occurring during the electrochemical gating process: 1) electric field-induced oxygen vacancy formation; and 2) ionic migration in both the channel material and ionic liquid. It is given by

$$i(t) = \frac{nFAD_O^{1/2}C_O^*}{\pi^{1/2}t^{1/2}}$$  (1)

where $i$ is the ionic current, $n$ is the number of electrons involved in the reaction, $F$ is the Faraday Constant, $A$ is the electrode area, $D_O$ is the diffusion constant of species O, $C_O^*$ is the initial concentration of species O, and $t$ is time. According to the Cottrell equation, the increasing rate of divalent nickel species is a nonlinear function of gating time duration. The Butler-Volmer equation with asymmetric electron transfer is applied to describe the influence of gate bias on regulating the changing rate of the divalent species. It is given by

$$i = i_0\left[e^{ne\eta/k_B T} - 1\right]$$  (2)

where $i_0$ is the exchange current, $\eta$ is the applied overpotential, $k_B$ is the Boltzmann constant, and $T$ is temperature. The Butler-Volmer equation shows that the rate is exponentially dependent on the gate bias. The application of both the Cottrell and Butler-Volmer equations allows for formulating the changing rate of nickel species. This approach is commonly used to study charge transfer processes in electrochemical reactions at interfaces. A detailed account of the formulation of the equations and their applications in electrochemical phenomena and devices can be found elsewhere.[31,32]

The change in conductivity of SNO can be estimated using the calculated divalent nickel species concentration from a percolation model.[30] As we are starting with near-stoichiometric films with high conductivity, we use the percolation model in the high density limit given by

$$\sigma(p) \sim \sigma_m(p - p_c)^l$$  (3)

where $\sigma_m$ is the conductivity in the low resistance state of SNO, $p$ is the volume fraction of conductive regions to insulating regions, $p_c$ is the percolation threshold, and $l$ is the critical exponent that determines the percolative behavior and is approximately 2 in three-dimensional systems. Here, divalent nickel regions are considered insulating, and trivalent nickel regions are considered conductive, typical of nickel oxides. The ratio of insulator to metal matrices can be simulated by the ratio of divalent nickel to trivalent nickel matrices. Divalent nickel species concentration can be approximated as discussed above. This leads to an expression for the voltage- and time-dependent resistance of the SNO device as

$$\frac{\sigma(V,t)}{\sigma_{\mathrm{m}}} = \left(A \mp B(e^{\pm C(V+V_{\mathrm{o}})} - 1)t^{1/2} - p_{\mathrm{c}}\right)^2 \frac{1}{(1-p_{\mathrm{c}})^2} \tag{4}$$

where $A$, $B$, $C$, and $V_{\mathrm{o}}$ are fitting parameters. $A \leq 1$ is a measure of the conductive region volume fraction before gating; $B$ is a combination of several constants including the prefactor of the integrated Cottrell equation (to obtain number density), exchange current, and normalization factors; $C$ is related to the number of electrons in the reaction and $V_{\mathrm{o}}$ is the overpotential at zero external bias. We use the common percolation threshold of $p_{\mathrm{c}} = 0.5$. The top (bottom) operator in the $\mp/\pm$ operators is used for positive (negative) gate bias. Several important effects due to electrode geometry, temperature, multiple defect types, gating history, and catalyst formation are difficult to quantify and are not considered here. The model should therefore be regarded as qualitative. Experimental data have been collected to characterize the relevant parameters, and sample fits for constant ±2.5 V gate voltage are shown in Figure 3.

## V.    Simulated neural mechanisms

We implement the model for potentiation and depression in our SNO devices into a custom MATLAB/Simscape component, which is in turn integrated into the circuit shown in Figure A1 for system simulations. The custom SNO component has three user-specified parameters: the minimum source-drain resistance, the initial source-drain resistance (which is akin to *a priori* experimentally gating the SNO to the neutral state), and the source-gate resistance (~20 MΩ). SNO1 is initially set to a low resistance level (800 Ω), corresponding to the US, and SNO2 is initially set to a moderate resistance level of 2 kΩ, corresponding to the NS. The output signals from N1 and N2 are shown in the top two panels of Figure 4a. The voltage sequence is similar to that in Figure 2. The simulation environment allows for direct reading of the time-dependent resistances of SNO1 and SNO2. The output of N3 is shown in the third panel of Figure 4a, and the simulated resistances of SNO1 and SNO2 are shown in the bottom panel of Figure 4a. It is clear that the circuit simulates classical conditioning similarly as the experimental demonstration in Figure 2a.

To simulate unlearning, we set the initial resistance of SNO2 to the conditioned level of 800 Ω and we reconfigure N1 as an inhibitory neuron that transmits negative forward propagating voltages (see Appendix A). Now, during the phase in which both N1 and N2 fire concurrently, N1 suppresses the excitatory signal from N2, and N3 does not fire. Thus, although SNO2 is conditioned, the signal from N2 does not trigger N3. The circuit responds by causing depression (resistance increase) of SNO2, as shown in Figure 4b. This process can be associated with the unlearning or counter Hebbian process in biological systems. After sufficient depression, the resistance of N2 becomes appreciably large such that N2 no longer triggers N3, and the CS-CR association is destroyed.

We can obtain similar classical conditioning with Hebbian learning between two *inhibitory* presynaptic neurons connected to an excitatory postsynaptic neuron using a modified synapse logic block while keeping the remainder of the circuit unchanged from above. The modified synapse logic block is shown in Figure A2 and is described in Appendix A. The block is a modified pass transistor logic AND gate. It maintains the property of not requiring an external power source, similar to the logic block of Figure A1b. With this synapse logic block, we can simulate identical conditioning behavior as shown in Figure 4a but with inhibitory presynaptic neurons (Figure 5a).

Aside from associative learning mechanisms, it is also useful to demonstrate non-associative mechanisms such as sensitization and habituation at the electronic device level. Sensitization is the process by which repetition of a stimulus leads to enhanced response to that stimulus, and habituation is the converse process. As discussed above, the circuit of Figure A1 has *a priori* sensitization-like functionality. A presynaptic signal that triggers a postsynaptic signal will cause LTP of the synaptic SNO interconnection until the SNO device reaches the minimum resistance level. With the same logic block of Figure A1, we can modify the circuit slightly to exhibit habituation as well. Simply by removing the back-propagating connection from N3 to the logic block of Synapse1, for example, we cause a situation where now there is no correlation between pre- and postsynaptic signals. In this case, each time N1 fires, a positive voltage is applied to the gate of SNO1 (causing LTD) because there is no back-propagating signal with which temporal overlap may occur. We simulate habituation with this circuit modification (Figure 5b). The resistances of SNO1 and SNO2 are initially both set to low values, and the input to N2 is grounded so that only N1 is active. With each pulse from the output of N1, the resistance of SNO1 increases slightly. Eventually, the resistance is sufficiently large as to no longer transmit an input voltage to N3 that is above threshold, and N3 is no longer triggered by N1. Therefore, not only can the circuit of Figure A1 be used for classical conditioning, sensitization, and unlearning, but a simple modification can extend the circuit capabilities to include habituation. In certain biological systems, synapses can exhibit both habituation and sensitization even though they are opposing processes, although the latter may occur only through interactions with neighboring synapses and neurons.[33] This is more of system-level learning rather than the device-level learning we are demonstrating here.

Aside from a variety of neuronal learning/unlearning mechanisms, we can implement the model to construct a more complex neural network based on feedback and recurrent excitation (Figure 6a) for memory storage. This network is inspired by the CA3 region in the hippocampus.[6] We use a modified synapse logic block that has only LTP functionality and not LTD. There are 5 input neurons (IN$i$) connected to 5 output pyramidal neurons (ON$j$). Each of the input neurons can fire or idle. Here IN1 and IN3 are firing, and the rest are silent (Figure 6b left). In this network, the direct synaptic connections (hatched boxes) between the input and output neurons are sufficiently strong such that firing of an input neuron always activates the corresponding output neuron. Overall, this topology is a feedforward excitation circuit. The important aspect is the recurrent excitation: every postsynaptic pyramidal output neuron is connected through axon collaterals (dashed lines) to all other output neurons, thus creating feedback. This means that every output neuron receives convergent information from all the other neurons in the network. We denote each synapse by the two neurons it connects, forming a connectivity matrix (S($i,j$) is the synapse between IN$i$ and ON$j$). Now letting the appropriate input neurons fire, and starting with low connectivity synapses (high resistances), only the synapses connected between two firing neurons (input and output) will strengthen through Hebbian-like learning, thus forming a true connectivity matrix between the firing pattern of the input and the output neurons. Results of the simulation can be seen in the right panel of Figure 6b, showing successful modification of only the appropriate synapses S(1,1), S(3,1), S(1,3), and S(3,3). The initial firing pattern is effectively stored in this connectivity matrix. The original input value can be retrieved by firing all input neurons over short time scales relative to the potentiation process. In this case the appropriate output neurons (ON1 and ON3) will exhibit higher values, since they receive not only the original input, but also the feedback from other firing neurons.

## Conclusions

In this work, we implement a model strongly correlated complex oxide samarium nickelate as a synapse-like device component in a brain-emulating neuronal circuit. Three-terminal $SmNiO_3$ devices in such networks are compatible with both excitatory and inhibitory neurons for exhibiting device-level classical conditioning, unlearning, and sensitization without circuit modifications. The $SmNiO_3$ devices can be modeled using an electrochemical ionic diffusion process under the conditions studied, and with the model, we simulate classical conditioning with excitatory and inhibitory neurons, associative unlearning, habituation, and a memory storage network. With the expansive ability to exhibit associative and non-associative learning modes with excitatory and inhibitory neurons, these studies may ultimately enable system-level neural simulation in electronic circuits, such as for investigating fear extinction,[34] neurodegenerative disease,[35] or perceptual learning.[36]

## Acknowledgements

This work was performed in part at the Center for Nanoscale Systems (NSF ECS-0335765), part of Harvard University. The authors acknowledge National Academy of Sciences and ARO-MURI (W911-NF-09-1-0398) for financial support. Y. M. is an awardee of the Weizmann Institute of Science – National Postdoctoral Award Program for Advancing Women in Science.

**Appendix A: Classical conditioning and unlearning circuit**

The full diagram for the circuit implemented in this work is shown in Figure A1a. Timed voltage sources are used as input stimuli for N1 and N2. The signals from N1 and N2 transmit through SNO1 and SNO2 as well as to the logic blocks. The signals from SNO1 and SNO2 then become inputs to N3, which sends back propagating signals from NoutBP to the logic blocks associated with SNO1 and SNO2. We use the excitatory neuron convention put forth by Pershin and Di Ventra wherein forward-propagating signals are positive voltage and back-propagating signals are negative voltage.[25] The presynaptic signals and SNO devices behave as a weighted-averaging circuit such that the input to N3 is given by $V_{N3,in} = (V_{N1,out}/R_{SNO1} + V_{N2,out}/R_{SNO2})/(1/R_{SNO1} + 1/R_{SNO2})$, according to Millman's Theorem. Here we use $V_{N1,out} = V_{N2,out} = V_N$. To illustrate the circuit behavior, let us assume that before conditioning we have $R_{SNO2} \gg R_{SNO1}$. Thus, when only N1 fires, $V_{N3,in}(N1) = V_N/(1 + R_{SNO1}/R_{SNO2}) \approx V_N$, and when only N2 fires, $V_{N3,in}(N2) = V_N/(1 + R_{SNO2}/R_{SNO1}) \approx 0$. After conditioning, $R_{SNO1} \approx R_{SNO2}$ and $V_{N3,in}(N1) \approx V_{N3,in}(N2) \approx V_N/2$. The threshold voltage for N3 to fire is therefore set slightly below $V_N/2$.

The neuron blocks (Figure A1c) are composed of two stages. In the first stage, a comparator compares the above weighted average to the threshold voltage connected to the non-inverting input of the comparator. If the input is above threshold, the comparator outputs a specified positive voltage, otherwise it outputs 0 V. The output of the comparator is taken as the forward propagating output of the neuron. The second stage of the neuron block is an op-amp wired as a unity-gain inverting amplifier that reverses the signal polarity of the comparator output. This is taken as the back propagating negative voltage from the neuron block. Because the stable voltage window for efficient synaptic modification of the SNO devices is between -2.5 V and 2.5 V, we use ±2.0 V voltage pulses as the outputs from the comparator and inverting amplifier to remain comfortably within the voltage window. An inhibitory neuron block can be implemented simply be taking the output of the inverting amplifier as the forward propagating signal and the output of the comparator as the back propagating signal.

The circuit diagram for the synapse blocks is illustrated in Figure A1b. The block is composed of two parts, a resistor voltage divider (bottom) and a logic component (top). The signal from the presynaptic neuron is the input to the voltage divider, and the output of the voltage divider is connected to the SNO device source terminal. The purpose of the voltage divider is such that a small but non-zero fraction of the presynaptic voltage is applied to the SNO source. This fraction is sufficiently large as to be readable by the comparator of the postsynaptic neuron block, but it is sufficiently small such that the source terminal of the SNO device remains near ground. For a presynaptic voltage of ±2.0 V, the voltage divider applies only ±25 mV to the SNO source. This allows us to use the positive and negative voltages from the pre- and postsynaptic neurons directly as gate voltages to achieve Hebbian-like LTD and LTP, respectively. If instead the full +2.0 V was applied to the SNO source from the presynaptic neuron, then +4.0 V would need to be applied to the gate to achieve a net gate-source voltage of +2.0 V for the LTD we demonstrate here. However, a +4.0 V supply cannot be implemented in our circuit while maintaining behavior analogous to biological systems such as passive synaptic circuitry. The logic component consists of two interconnected n-channel MOSFETs, one depletion-mode, one enhancement-mode, and both with $|V_{th}| \sim 3$ V. Connected in this manner, the logic component outputs the truth table shown in the inset of Figure A1b to the gate of the SNO device. LTP occurs only if pre- and postsynaptic signals temporally overlap, indicating an association between the firing of the interconnected neurons. If the presynaptic neuron fires but the postsynaptic neuron does not, then there is no association and LTD occurs. This is similar to Hebbian learning in biological neural systems.

The synapse logic block for classical conditioning between two inhibitory neurons connected to an excitatory neuron is shown in Figure A2. The inhibitory neuron transmits a negative voltage signal, the excitatory postsynaptic neuron transmits a negative back-propagating voltage signal, and the voltage that needs to be applied to the SNO gate for potentiation is also negative. Therefore, the circuit needed to achieve potentiation only for concurrent pre- and postsynaptic signals is equivalent to a logic AND gate in which logic level 1 is -2 V. As with the block of Figure A1b, the circuit should not require external power for the synaptic block to remain overall passive, as in biological synapses. The circuit of Figure A2 is a modified pass transistor logic AND gate that is suitable for demonstrating classical conditioning with inhibitory presynaptic neurons. It is composed of one n-channel and one p-channel MOSFET, both enhancement-mode with $|V_{th}| < 2$ V.

**Figure Legends**

Figure 1:  **a**, Schematic of neural classical conditioning circuit. Circles represent neurons and triangles represent synapse connections and neuron outputs. Neurons N1 and N2 accept US and NS/CS, respectively. **b**, Schematic of electronic classical conditioning/unlearning circuit used in this work. US and NS/CS signals transmit through SNO1 and SNO2 to neuron N3. Back-propagating signal (dashed line) from N3 correlates with signals from N1 and N2 at logic blocks, which apply voltage to gate of SNO1 and SNO2 for potentiation or depression, if necessary. **c**, Illustration and optical micrograph of 3-terminal SNO synaptic device. Illustration shows ionic liquid (IL) interfaced with SNO channel along with source (S), drain (D), and gate (G) electrode labels. SNO substrate is LaAlO$_3$.

Figure 2:    **a**, Classical conditioning experimental results. N1 and N2 inputs are positive excitations representing US and NS/CS, respectively. Initially, SNO1 is in a low resistance state and SNO2 is in a high resistance state. After the conditioning phase, both resistances are low and N3 fires when either N1 or N2 is triggered. **b**, Unlearning experimental results using same circuit as for part (a). N1 is reconfigured as an inhibitory neuron, which accepts a positive input stimulus but outputs a negative signal. Initially, both SNO1 and SNO2 have low resistances. After the unlearning phase, the resistance of SNO2 is increased such that triggering N2 no longer causes N3 to fire.

Figure 3:    Representative time-dependent conductivity of SNO synaptic device for -2.5 V (solid shapes, left axis) and +2.5 V (hollow shapes, right axis) gate voltage applied and held at $t = 0$, and fit to data using Eq. 4. Experimental data are in black and results of model fit are in red.

Figure 4:    **a**, Simulated classical conditioning with excitatory input neurons and excitatory output neuron using device model. SNO1 is the unconditioned synapse and SNO2 is initially the neutral synapse. The range $t$ = 0-60 s is the initial probing phase. $t$ = 60-300 s is the conditioning phase, which shows a clear monotonic decrease in SNO2 resistance with time. $t$ = 300-350 s is the final probing phase in which an output from N3 is triggered when either N1 or N2 fires, indicating the occurrence of conditioning. **b**, Unlearning simulation with same voltage sequence as **a** but with N1 configured as an inhibitory neuron. SNO2 is initially set to the low resistance state achieved after the conditioning of **a**. Here, firing of N3 is suppressed by the inhibitory signal of N1, causing resistance increase of SNO2 with time. In the final probing phase, N3 is no longer triggered from N2 firing.

Figure 5:  **a**, Classical conditioning simulation results using logic block of Figure A2 with two inhibitory input neurons. N3 is configured here to fire for negative inputs below threshold. Classical conditioning of SNO2 clearly occurs similarly as with excitatory input neurons in Figure 4a. **b**, Habituation simulation with excitatory input and output neurons by disconnecting back-propagating signal line of Figure 1b. When N1 fires, a back-propagating signal is not received at the logic block, and the SNO1 resistance correspondingly increases even though N3 is initially firing in response to N1. Eventually, SNO1 resistance increases sufficiently such that N3 no longer fires in response to N1, similar to habituation.

Figure 6:  **a**, Schematic of feedforward recurrent memory network composed of SNO synaptic devices. Input neurons and signals (red) are the presynaptic connections to the synapses (gray). Output signals from the synapses sum at the output neurons (blue),

and back propagating signals (dashed lines) feed back to the synapses. Firing neurons (solid circles) cause certain synaptic resistances (solid rectangles) to decrease, while the remaining neurons and synapses remain idle (unfilled circles/rectangles). The hatched rectangles represent direct, permanent low resistance connections between input and output neurons for the feedforward functionality. **b**, Input neuron signals (left) and corresponding synaptic response (right). Only synapses related to the input firing pattern exhibit enhanced weight and decreased resistance.

Figure A1: **a**, Overview classical conditioning/unlearning circuit schematic composed of excitatory input neurons (red), synapse logic blocks (orange), SNO devices (yellow), and excitatory output neuron (blue). **b**, Schematic of synapse logic block with accompanying truth table. Voltage divider is used to apply small but non-zero voltage to SNO device such that $V_G \approx V_{GS}$ and signals from pre- or postsynaptic neurons can be used directly for depression/potentiation. **c**, Schematic of neuron blocks. First comparator stage compares input to specified threshold value and outputs a positive voltage signal. Second op-amp stage inverts polarity of comparator output for negative back propagating signal.

Figure A2: Synapse logic block schematic that is equivalent to AND gate with -2 V as logic level 1. Voltage divider has same functionality as in Figure A1b.

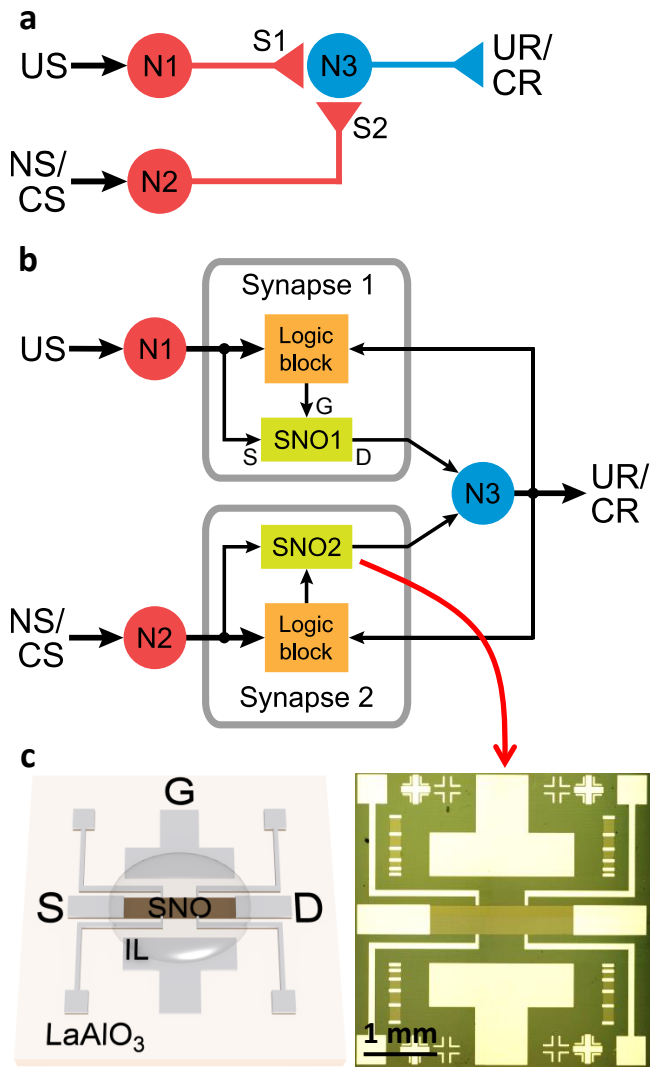

Figure 1

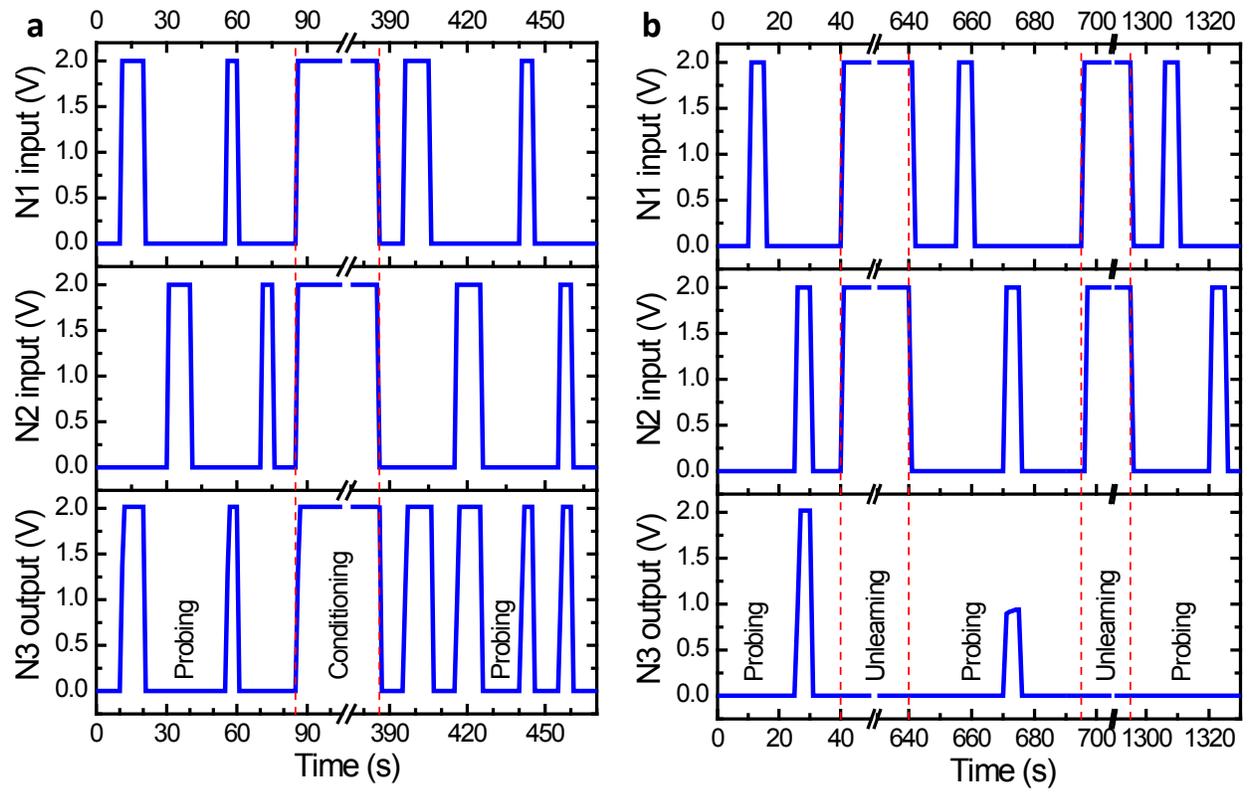

Figure 2

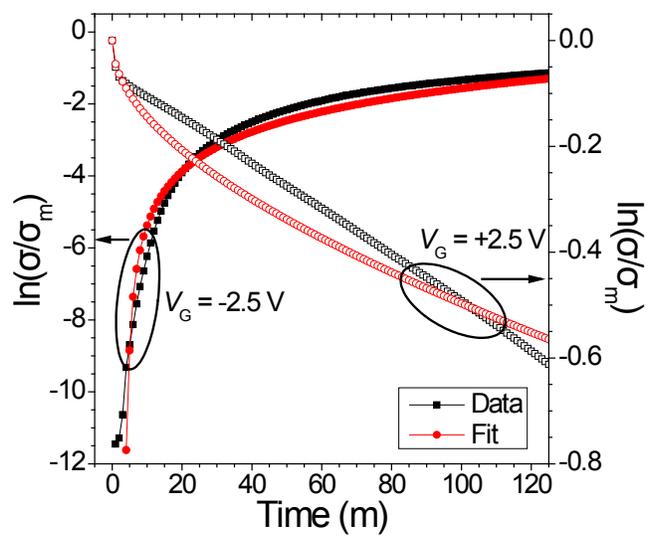

Figure 3

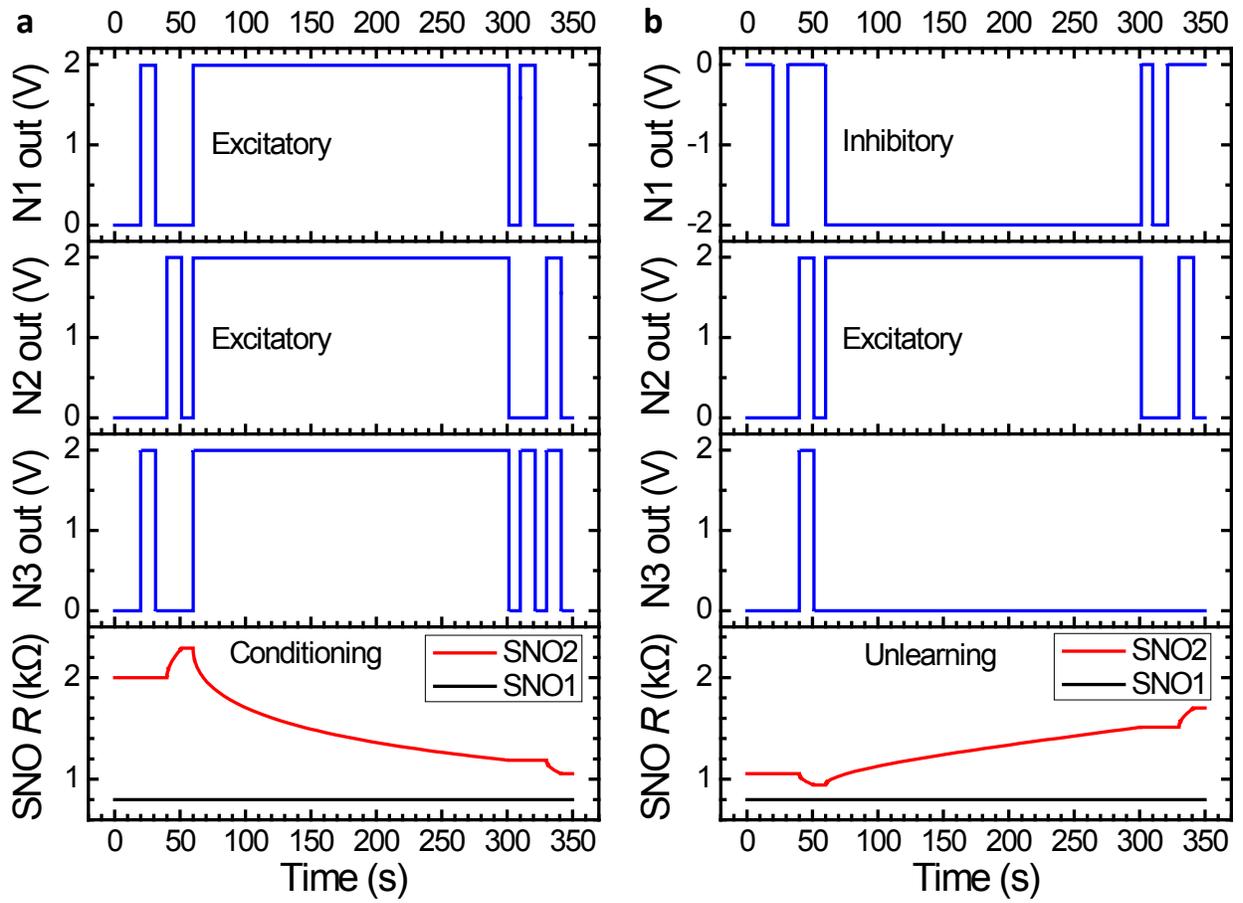

Figure 4

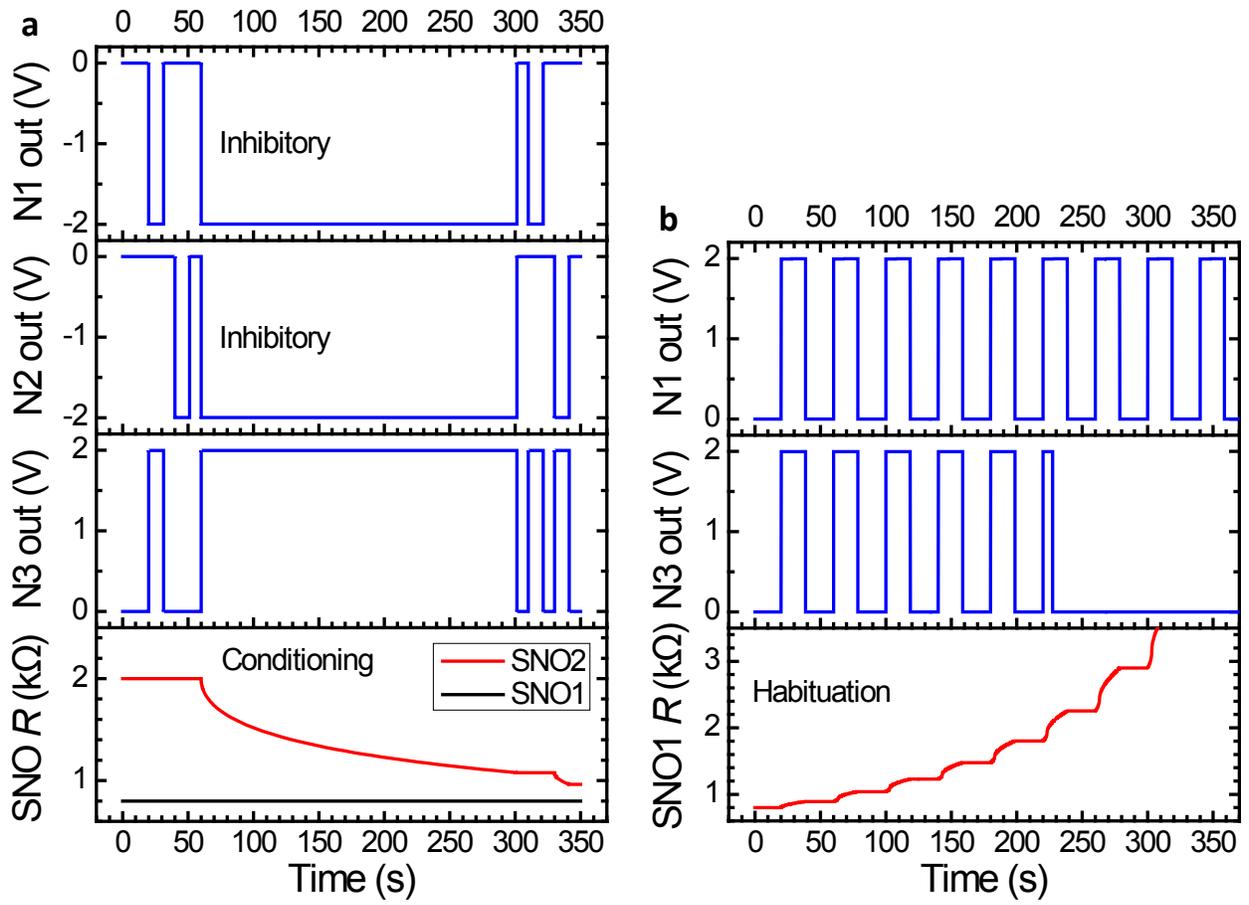

Figure 5

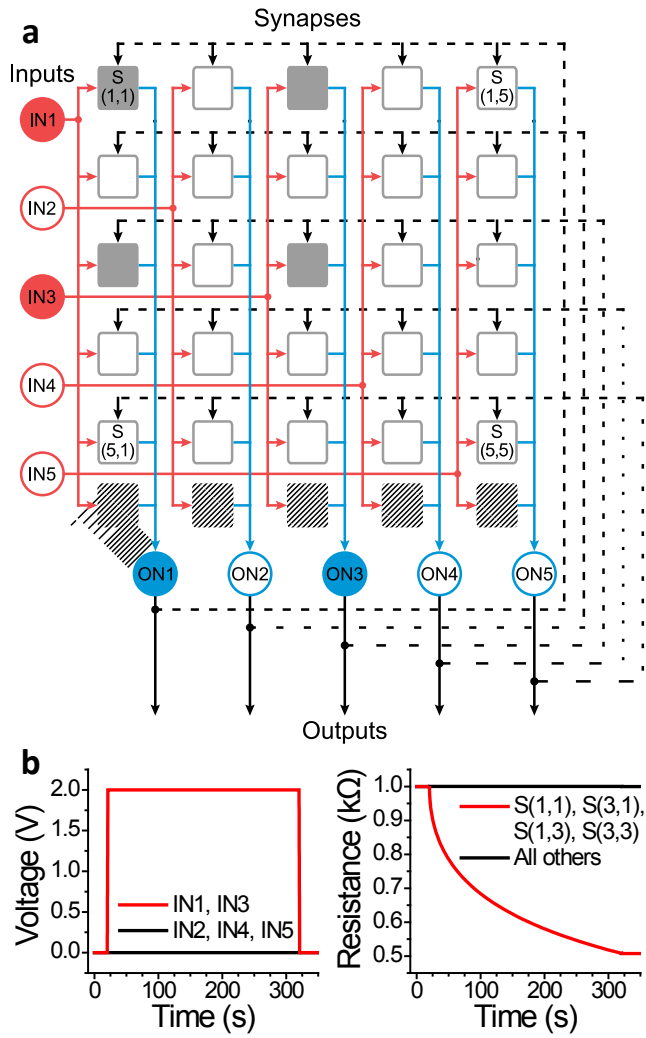

**a**

Synapses

Inputs

S(1,1) ... S(1,5)

IN1
IN2
IN3
IN4
IN5

S(5,1) ... S(5,5)

ON1  ON2  ON3  ON4  ON5

Outputs

**b**

Voltage (V) vs Time (s):
— IN1, IN3
— IN2, IN4, IN5

Resistance (kΩ) vs Time (s):
— S(1,1), S(3,1), S(1,3), S(3,3)
— All others

Figure 6

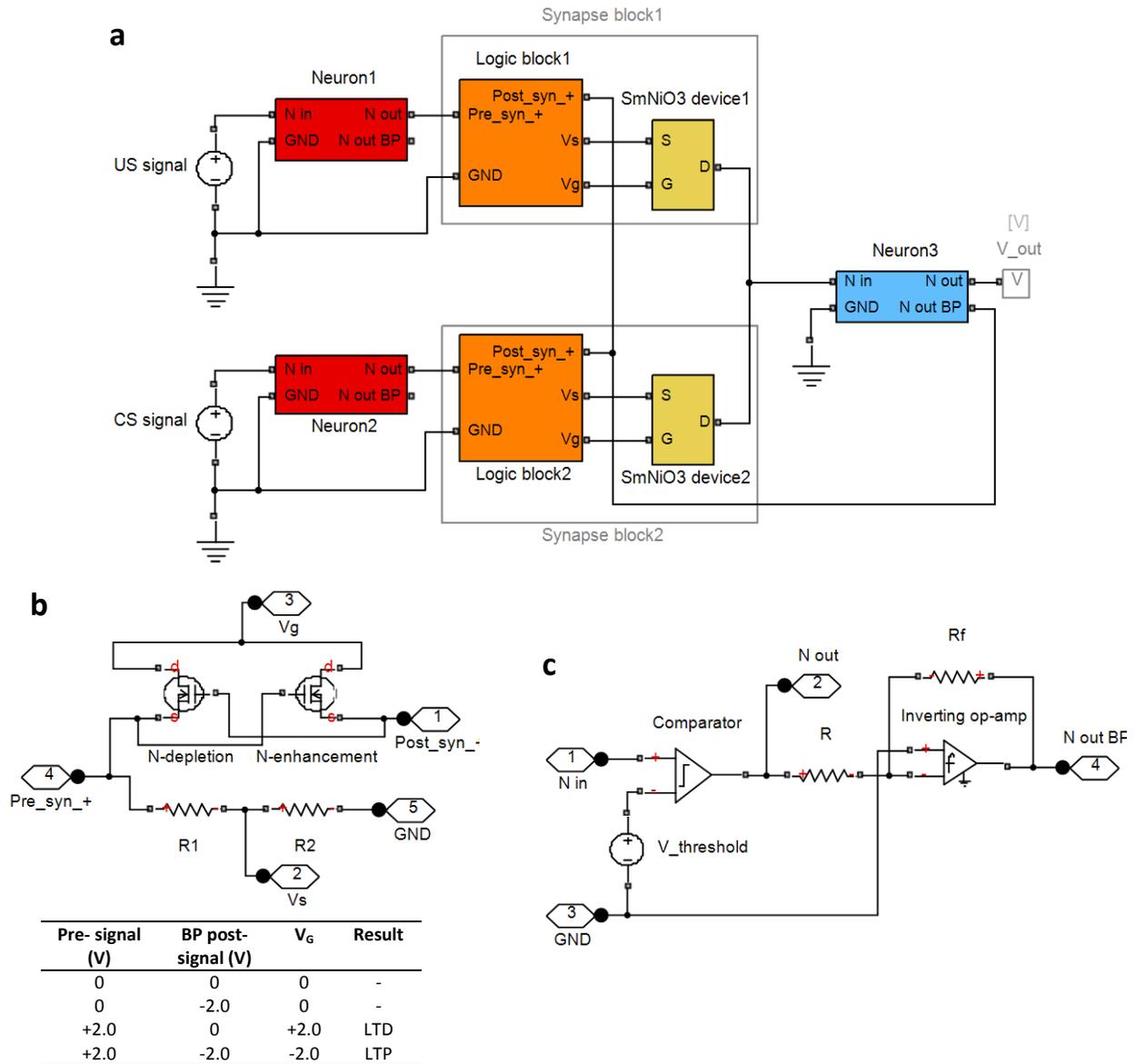

| Pre- signal (V) | BP post-signal (V) | $V_G$ | Result |
|---|---|---|---|
| 0 | 0 | 0 | - |
| 0 | -2.0 | 0 | - |
| +2.0 | 0 | +2.0 | LTD |
| +2.0 | -2.0 | -2.0 | LTP |

Figure A1

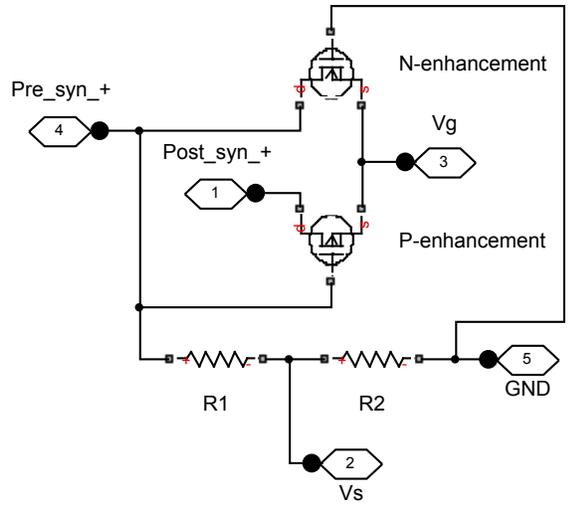

Figure A2